\documentclass[12pt]{article}
\usepackage{epsf}

\newcommand{\eeq}{\end{equation}}

\newcommand{\beq}{\begin{equation}}
\newcommand{\beql}{\begin{eqnarray}}
\newcommand{\eeql}{\end{eqnarray}}

\begin{document}

\title{ Parity Effects in Eigenvalue Correlators, 
Parametric and Crossover Correlators in Random Matrix Models:
Application to Mesoscopic systems }

\author{N. Deo\\
Poornaprajna Institute of Scientific Research,\\
Sadashiva Nagar, Bangalore 560080, India}

\maketitle

\begin{abstract}

This paper summarizes some work 
I've been doing on eigenvalue correlators of Random Matrix 
Models which show some interesting behaviour.
First we consider matrix models with gaps in there spectrum or 
density of eigenvalues.
The density-density correlators of these models  
depend on whether N, where N is the size of the matrix, 
takes even or odd values. The fact that this dependence persists 
in the large N thermodynamic limit is an unusual property and may have 
consequences in the study of one electron effects in mesoscopic
systems. Secondly, we study the parametric
and cross correlators of the Harish Chandra-Itzykson-Zuber 
matrix model. The analytic 
expressions determine how the correlators change as a parameter
(e.g. the strength of a perturbation in the hamiltonian of the 
chaotic system or external magnetic field on a sample of material)
is varied. The results are relevant for the conductance fluctuations 
in disordered mesoscopic systems.   

\end{abstract}



%\submitted{{\noindent \it }}
\section{Introduction}\label{intro}

Recently there is a lot of activity in the field of quantum chaos and 
mesoscopic systems. This has resulted in the study of eigenvalue correlators
in large random matrix models ref. \cite{mg}-\cite{jain}. 
This paper summarizes some work I've been
doing on calculating eigenvalue correlators in random matrix models 
which show novel properties which maybe useful in the study of transport 
properties in mesoscopic conductors. The first models are random matrix 
models with gaps in the eigenvalue spectrum
and the second are two-matrix models first considered 
by Harish Chandra-Itzykson-Zuber ref. \cite{ciz}. 
In the first model the `fine grained' 
correlators are found using the method of orthogonal polynomials 
ref.\cite{d97}. These
correlators are unusual in that in the large N thermodynamic limit they
tend to different limits depending on whether N goes to infinity through
even or odd ref. \cite{bd}. This property maybe found in mesoscopic systems 
which are sensitive to single electron effects. 
The second models are the two-matrix models of Harish Chandra-
Itzykson-Zuber ref. \cite{djs,d96}. Here correlators are calculated 
using the Dyson-Schwinger
method and give the long ranged parametric and cross-correlators. Transport 
experiments involving changes in magnetic fields in which long range 
eigenvalue statistics are effected will be a testing ground for these 
results.

I shall first discuss the double well matrix model 
with potential $V(M)= -{\mu\over 2} M^2 + {g\over 4} M^4$ where 
$M$ is a random $N\times N$ matrix. I will present some results
for the two-point density-density correlation functions which
show interesting parity effects and further characterizes these 
models in a new universality class. The original 
model is $Z_2$ symmetric while there is $Z_2$ symmetry breaking 
in the correlation functions. Second, the Chandra-Itzykson-Zuber
matrix model is discussed and the results for the smoothed
long range parametric and crossover eigenvalue correlators, 
found using the Dyson-Schwinger equations, are given.
 
\section{Notations and Conventions for the Double-Well Matrix Model}
\label{note}

We start by establishing the notations and conventions. Let
$M$ be a hermitian matrix. The partition function to be
considered is $ Z=\int dM e^{-N tr V(M)} $
where $M=N\times N$ hermitian matrix. The Haar measure 
$dM = \prod_{i=1}^{N}dM_{ii}\prod_{i<j} dM_{ij}^{(1)}
dM_{ij}^{(2)}$ with $M_{ij}=M_{ij}^{(1)} + i M_{ij}^{(2)}$ 
and $N^2$ independent variables. $V(M)$ is a polynomial in M:
$
V(M)=g_1 M + (g_2/2) M^2 + (g_3/3) M^3 + (g_4/4) M^4 + .....
$
The partition function is invariant under the change of variable
$M^{\prime}=U M U^{\dagger}$ where $U$ is a unitary matrix. We can 
use this invariance and go to the diagonal basis ie 
$D^{\prime}=U M U^{\dagger}$ such that $D^{\prime}$ is the matrix
diagonal to $M$ with eigenvalues $\lambda_1,\lambda_2,.....\lambda_N$.
Then the partition function becomes
$
Z = C \int_{-\infty}^{\infty} \prod_{i=1}^{N} d\lambda_i 
\Delta (\lambda)^2 e^{-N \sum_{i=1}^N V(\lambda_i)}
$
where $\Delta (\lambda) = \prod_{i<j} |\lambda_i-\lambda_j|$
is the Vandermonde determinant. The integration over the group
U with the appropriate measure is trivial and is just the constant C. 
By exponentiating the determinant as a `trace log' we arrive at the 
Dyson Gas or Coulomb Gas picture. The partition function is simply
$
Z = C \int_{-\infty}^{\infty} \prod_{i=1}^N d\lambda_i e^{-S(\lambda)} 
$
with
$ S(\lambda) = N \sum_{i=1}^N V(\lambda_i) - 2
\sum_{i,j,i\ne j} ln |\lambda_i-\lambda_j|$. 

\begin{figure} 

\leavevmode
\epsfxsize=4in
\epsffile{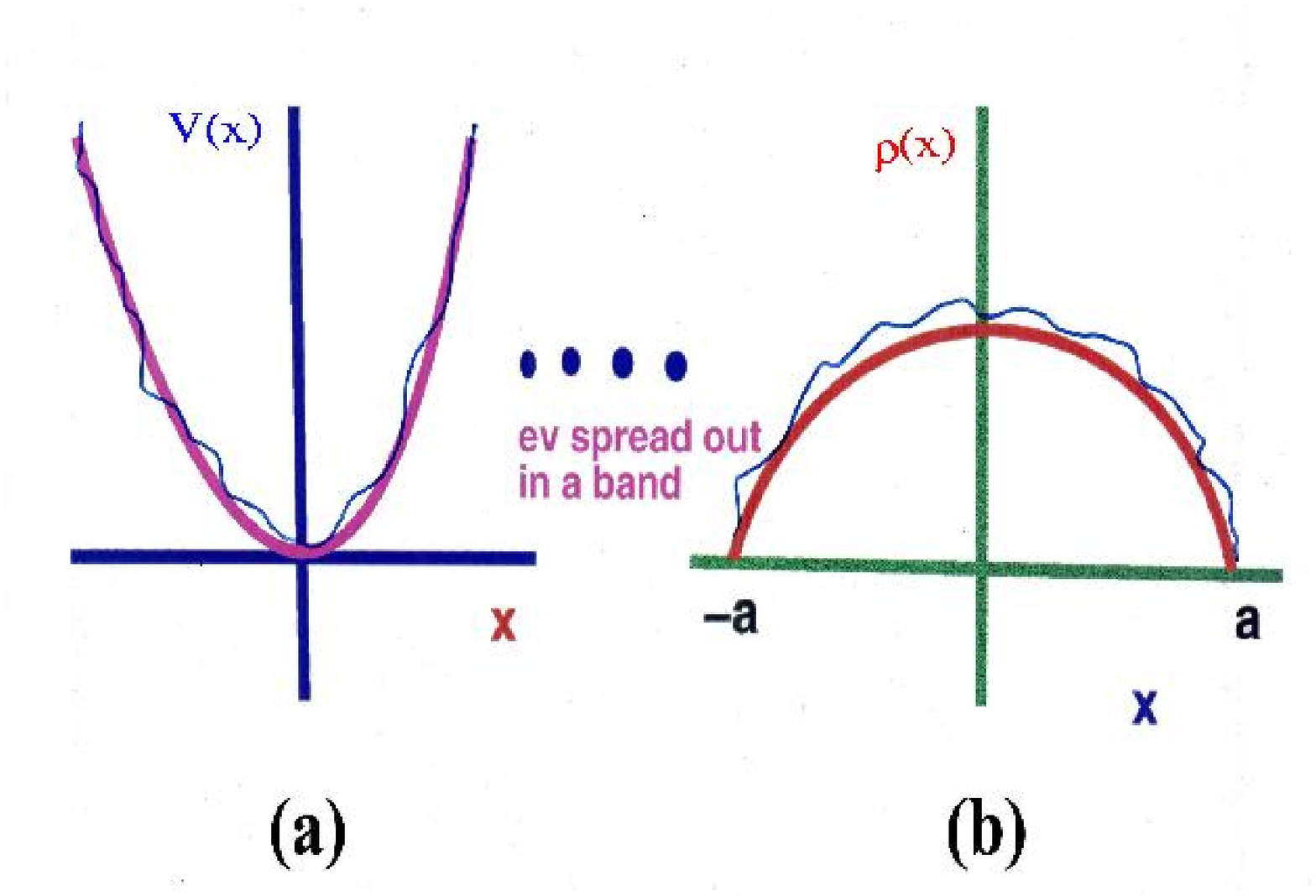}

%\vspace{6cm}
%\special{psfile=Fig1ab.eps hscale=100 vscale=100 hoffset=0 voffset=0}
%\begin{center}
%\scalebox{0.3}{\includegraphics{Fig1ab.eps}}
%\end{center}
%\hbox to\hsize{\epsfxsize=0.8\hsize\hfil\epsfbox{Fig1ab.eps}
%\hfil}

\caption{ (a). The confining potential (b). The density of eigenvalues }
\label{fig1ab} 
\end{figure}

This is just a system of N particles with coordinates $\lambda_i$ on
the real line, confined by a potential and repelling each other with 
a logarithmic repulsion. The
spectrum or the density of eigenvalues $\rho (x) =
{1\over N} \sum_{i=1}^N \delta (x-\lambda_i)$ is in the large N limit
or doing the saddle point analysis just the Wigner semi-circle for a
(Gaussian probability distribution for the eigenvalues) quadratic
potential. The physical picture is that the eigenvalues try to be at the 
bottom of the well. But it costs energy to sit on top of each other
because of logarithmic repulsion, so they spread. $\rho$ has support on a 
finite line segment. This continues to be true whether the potential is
quadratic or a more general polynomial and only depends on there being
a single well though the shape of the Wigner semi-circle is correspondingly 
modified. For the quadratic potential the density is 
$\rho (x) = {1\over \pi} \sqrt{(x-a)(b-x)}$ where $ [a,b] $ are 
the end of the cuts. See Fig. \ref{fig1ab}.

On changing the potential more drastically by having two humps or wells,
the simplest example being a potential 
$V (M) = - {\mu\over 2} M^2 + {g\over 4} M^4$, 
the density can get disconnected support. The precise expressions for the 
density of eigenvalues are as follows:

\beql
\rho (x) &=& {g\over \pi} x \sqrt{(x^2-a^2)(b^2-x^2)}  ~~~~~~~~   
 a<x<b \nonumber\\   
&=& 0 ~~~~~~~~ -b<x<-a 
\eeql

where $ a^2 = {1\over g} [|\mu|-2\sqrt{g}]$ and
$ b^2 = {1\over g} [|\mu|+2\sqrt{g}]$
and $|\mu|>2\sqrt{g}$, which is the condition that the wells 
are sufficiently deep. The eigenvalues sit in symmetric bands
centered around each well. Thus $\rho$ has support on two line
segments. As $|\mu|$ approaches $2\sqrt{g}$, $ a \rightarrow 0$ and
the two bands merge at the origin. The density is then
\beql
\rho (x) &=& {{g x^2}\over \pi} \sqrt{x^2-{2\mu\over g}}
~~~~~~~~ -\sqrt{2|\mu|\over g} < x < \sqrt{2|\mu|\over g} \nonumber \\
&=& 0 ~~~~~~~~ otherwise.
\eeql

\begin{figure} 

\leavevmode
\epsfxsize=4in
\epsffile{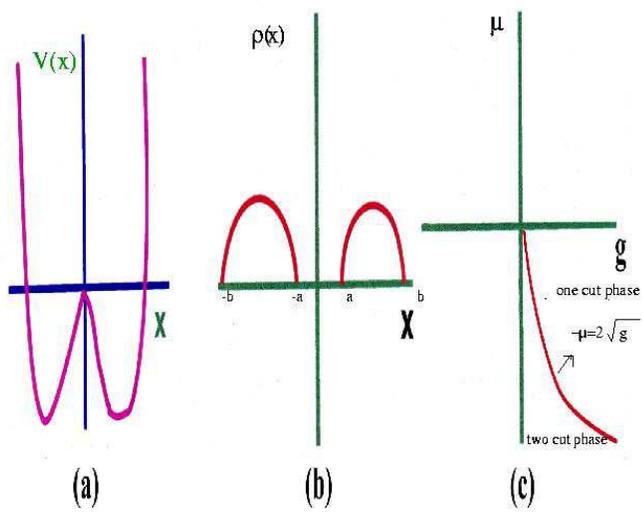}

%\vspace{6cm}
%\special{psfile=Fig2ab.eps hscale=100 vscale=100 hoffset=0 voffset=0}
%\begin{center}
%\scalebox{0.3}{\includegraphics{Fig2ab.eps}}
%\end{center}
%\hbox to\hsize{\epsfxsize=0.8\hsize\hfil\epsfbox{Fig2ab.eps}
%\hfil}
\caption{ (a). The double-well potential (b). Density of eigenvalues 
(c). The phase diagram }
\label{fig2ab} 
\end{figure}

The phase diagram and density of eigenvalues for the $M^4$ potential 
is shown in Figs. \ref{fig2ab}.

The simplest way to determine $\rho (z)$ explicitly is to use the
generating function $F(z)=<{1\over N} Tr {1\over {z-M}}>$ and
its saddle point or Schwinger-Dyson equation also known in the 
mathematics literature as the Riemann-Hilbert problem 
$ F(z) = {1\over 2} [ V^{\prime} (z) + \sqrt{\Delta} ] $
with $\Delta (z) = V^{\prime} (z)^2 - 4 b(z)$
and $b(z) = g z^2 + \mu + g < {1\over N} Tr M^2 >$ 
(see ref. \cite{bdjt}). The density
$\rho (x)$ is then determined by the formula 
$\rho (z) = -{1\over 2\pi} Im \sqrt{\Delta (z)}$. 

In what follows I will outline the recurrence coefficient method 
of the orthogonal polynomials for the two-cut matrix model see
ref. \cite{bdjt} for more details. I give the results for the 
two-point correlators (also known as the 
`smoothed' or `long range' correlators) for all three ensembles.
Then the two-matrix model of Chandra-Itzykson-Zuber is defined
and the expressions for the parametric and crossover correlators 
are given.

\section{Orthogonal Polynomial Approach}
\label{ortho}

The partition function $Z$, can be rewritten
in terms of the orthogonal polynomials $P_n$ where 
the polynomials are defined as
$
\int_{-\infty}^{\infty} d\lambda e^{-NV(\lambda)}
P_n (\lambda) P_m (\lambda) = h_n \delta_{nm}
$
then $P_n = \lambda^n + l.o.$ and $P_0 (\lambda)=1$,
$P_1 (\lambda) = \lambda + c$, $P_2 (\lambda) 
= \lambda^2 + a \lambda + b $ ...... Then $Z$ the
partition function in terms of the orthogonal polynomials 
is
\beq
Z = \int \prod_{i=1}^N d\lambda_i e^{-N \sum V(\lambda)}
\left|
\begin{array}{ccc}
P_0 (\lambda_1)&....&P_0 (\lambda_N)\\
P_1 (\lambda_1)&....&P_1 (\lambda_N)\\
.\\
.\\ 
.\\
P_{N-1} (\lambda_1)&......&P_{N-1} (\lambda_N)\\
\end{array}
\right|.
\eeq
The partition function $Z$ is also known if we know
the $h_n$'s as the partition function can be expressed
in terms of the $h_n$'s 
$
Z=N! h_0 h_1 h_2.....h_{N-1}.
$
For example: 
N=2 case
$
\int d\lambda_1 d\lambda_2 e^{-N V(\lambda_1)-N V(\lambda_2)}
(P_0 (\lambda_1) P_1(\lambda_2)-P_0 (\lambda_2) P_1 (\lambda_1))^2
= h_0 h_1 + h_0 h_1  
= 2! h_0 h_1.
$
So the question is how does one find the $h$'s? 

The $P_n$ satisfy recurrence relations
\beq
x P_n = P_{n+1} + S_n P_n + R_n P_{n-1}.
\label{rec}
\eeq
Note that 
$ \int x P_n P_{n-2} e^{-NV(x)} dx =0 $
as $ x P_{n-2} = P_{n-1} + l.o. $. Thus $P_{n-2}$ $\&$ l.o. terms
do not appear on the right hand side of the recurrence relation
eq. (\ref{rec}). Then as $h_n=h_{n-1}R_n$ the product
$
h_0 h_1 ...... h_{N-1} = 
h_0 (h_0 R_1) (h_0 R_1 R_2) \\
...... (h_0 R_1... R_N) 
= h_0^N R_1^{N-1} R_2^{N-2} ...... R_{N-1}.
$
The free energy  
$
\Gamma = \ln Z = \ln N! + N \ln h_0 + \sum_{n=1}^{N-1} (N-n) \ln R_n
$
hence we need recurrence coefficients $R_n$'s to get the free energy.   

\section{Asymptotic Ansatz for the Orthogonal Polynomials of the
Symmetric Double-Well Matrix Model}\label{asymp}

We have been able to derive            
in ref. \cite{d97} for the symmetric double-well matrix model
the asymptotic ansatz for the orthogonal polynomials 
$\psi_n (\lambda) = 
{P_n (\lambda) \over \sqrt{h_n}} exp(-{N\over 2} V(\lambda))$
which is
\beq
\psi_n (\lambda) = {1\over {\sqrt{f(\lambda)}}} 
[ \cos (N\zeta - (N-n)\phi + \chi + (-1)^n \eta) (\lambda) 
+ O ({1\over N}) ]
\eeq
where $f$, $\zeta$, $\phi$, $\chi$  and $\eta$ are functions of $\lambda$,
and $\psi_n$ is damped outside of the cuts. We show that
$
f(\lambda)={\pi\over 2\lambda} {(b^2-a^2)\over 2} 
\sin 2\phi(\lambda)
$  
from the orthogonality condition satisfied by the orthogonal polynomials.
$
\zeta^{\prime} (\lambda) = - \pi \rho (\lambda)
$
from the relation $K(\lambda,\lambda)=\rho(\lambda)$ where $K(\mu,\nu)$
is the kernel defined by 
${1\over N}\sum_{i=1}^{N-1} \psi_i (\mu) \psi_i (\nu)$ and determines all
eigenvalue correlators. While $\phi (\lambda)$ and $\eta (\lambda)$ are
determined from the recurrence relations satisfied by the orthogonal
polynomials
$
\cos 2\phi (\lambda) = { {\lambda^2-({a^2+b^2 \over 2})}\over 
{({b^2-a^2 \over 2})}},  
$
$
\cos 2\eta (\lambda) = {{b \cos \phi (\lambda)}\over \lambda},
$
and
$
\sin 2\eta (\lambda) = {{a \sin \phi (\lambda)}\over \lambda}.
$

$f$, $\phi$, $\eta$ and $\chi$ are universal functions independent of
the potential, the only dependence on V enters through the endpoints
of the cuts a and b. $\zeta$ is non-universal since the eigenvalue
density depends in general on the detailed form of V. More details
are to be found in the ref. \cite{d97,bd}. From this one can
establish that the gapped matrix model is in a new universality 
class. 

Using the asymptotic ansatz the full density-density correlation
function maybe found and is given in ref. \cite{d97}. 
A simpler result are the `smoothed' or `long range' 
two-point correlation functions 
found in ref. \cite{bd} using a method of steepest decent
which has a constant C which is undetermined unless one is
explicitly in the symmetric case where $C=(-1)^N$ ref. \cite{bd} 
(ref. \cite{kf} obtain this result using a method due to Shohat),
other values for C have been found earlier using the 
loop equation method ref. \cite{aa}. Recently in ref. \cite{david}
it was shown that for double wells with equal depths but unequal 
widths, in the limit of the symmetric double wells, give the value
$C=(-1)^N$ confirming the results of ref. \cite{bd}.

\beql
4 \pi^2 N^2 \rho^c_2 (\lambda,\mu) &=& 
{ {\epsilon_{\lambda}\epsilon_{\mu}} \over {\beta \sqrt{|\sigma(\lambda)|}
\sqrt{|\sigma(\mu)|}} } \nonumber \\
\left( { {\sigma(\lambda)+\sigma(\mu)} \over {(\lambda-\mu)^2} } 
+ { {\sigma^{\prime} (\lambda) - \sigma^{\prime} (\mu)}
\over {(\lambda-\mu)} } +  
\lambda^2 + \mu^2 -{s\over 2} (\lambda+\mu) + 2C \right).
\eeql

Here $\sigma(z)=(z^2-a^2)(z^2-b^2)$, $s=a_1+a_2+a_3+a_4$, 
$\epsilon_{\lambda}=+1~~ for~~ a_3<\lambda<a_4, \epsilon_{\lambda}=-1~~for~~
a_1<\lambda<a_2$ and $\beta=1,2,4$ depending on whether $M$ the matrix
is real orthogonal, hermitian or self-dual quartonian. This result is 
different for even and odd N and hence has broken the $Z_2$ symmetry. It
would be very interesting to be able to see this effect in experiments
of mesoscopic systems which are sensitive to single electron effects.

\section{The Harish Chandra-Itzykson-Zuber matrix model and density-density
correlators}

The two-matrix model of Harish Chandra-Itzykson-Zuber is defined by the
partition function
\beq
Z  =  \int dA dB e^{-S}
\label{partab}
\eeq
\noindent
where $S=NSp[V(A)+V(B)-cAB]$ and $V(A)={1\over 2} \mu A^2$. For 
$\lambda = {1\over 2}$, $A$ and $B$ are $N\times N$ real symmetric
matrices (orthogonal ensemble), for $\lambda=1$, $A$ and $B$ are 
$N\times N$ Hermitian matrices (unitary ensemble), and for $\lambda=2$
$N\times N$ real self-dual quaternions (symplectic ensemble). $Sp (A)$
stands for $ Tr A$ for $\lambda={1\over 2},1$ and for ${1\over 2} Tr A$
for $\lambda =2$. In this model the connected density-density correlator
is $ \rho_{AB} (x,y) \equiv <\hat{\rho}_A (x) \hat{\rho}_B (y)>_c $, 
where the density is defined as $\hat{\rho}_A (x) \equiv {1\over N}
 Tr \delta (x-A)$, $<X> \equiv {1\over Z} \int dA dB e^{-S} X$, and the
subscript $c$ implies connected part. In ref. \cite{djs} the Dyson-Schwinger
equations are used to derive these eigenvalue correlators. Here only the 
results and physical interpretations are elaborated on.

In the large-N limit, the expectation value for the density is given by
the well known Wigner semicircle law 
\beql
<\hat{\rho}_A(x)>=<\hat{\rho}_B(x)>={2\over \pi a^2} \sqrt{a^2-x^2}, 
|x|\leq a,
\eeql   
\noindent
and $<\hat{\rho}> = 0$ for $|x|\geq a$, where $a$, the ``end point of
the cut'' is given by $a= ({4\lambda\mu\over {\mu^2-c^2}})$. The result 
for the connected density-density correlator to leading order in 
${1\over N}$, valid over the entire cut, is
\beql
\rho_{AB} (x,y) = - {1\over 4 \pi^2 N^2} {1\over \lambda a^2} 
{1\over \cos \theta \cos \alpha} \times \nonumber\\
\left [ { {1+\cosh u 
\cos (\theta+\alpha)} \over { [\cosh u + \cos (\theta + \alpha)]^2} }
+ { {1-\cosh u \cos (\theta-\alpha)} \over 
{ [\cosh u - \cos (\theta - \alpha)]^2} } \right ],
\eeql  
\noindent
where $ u \equiv ln ({mu\over c})$, $|x|, |y| \le a$, and we have
defined $\sin \theta = {x\over a}$ and $\sin \alpha = {y\over a}$.
For $ \lambda = 1 $ the above result was derived in ref. \cite{bz}
using different methods from the Dyson-Schwinger method. The 
Dyson-Schwinger method is capable of generalization to 
$\lambda={1\over2},2$ with the above result. Then these results are 
relevant for the calculation of conductance fluctuation of mesoscopic 
systems in which the magnetic field is changing.  

When one is interested in transitions from one symmetry class to
another, a hamiltonian $A$ is considered consisting of two parts $B$
and $V$ each drawn from different ensembles $A=B+V$. The         
partition function $Z$ and action $S$ are modified to have
\beql
V(A)={1\over 2} \mu_1 A^2,~~~V(B)={1\over 2} \mu_2 B^2
\eeql
and
\beql
c=\mu_1,~~~ \mu_2={ {\mu_1(1-\mu_1N)}\over {(2-\mu_1N)}}.
\label{param}
\eeql
The constant $\mu_1$ measures the strength of the perturbation. At
$\mu_1=\infty$ we get GOE, while $\mu_1={2\over N}$ gives GUE. However,
we will work in the more general case where $\mu_1$, $\mu_2$ and $c$ 
are independent $O(1)$ parameters. This is just the standard two-matrix model
except that A and B are drawn from different ensembles. The choice of 
parameters eq. (\ref{param}), mentioned above corresponds to the crossover
from GOE to GUE is then a special case of the formula derived.

We are interested in calculating the connected density-density correlator
\beq
\rho^c_{AA} (x,y) \equiv <\hat{\rho}_A(x)\hat{\rho}_A(y)>_c
\eeq
The full smoothed global result for the connected eigenvalue correlator
is (see ref. \cite{d96})
\beq
\rho^c_{AA} (x,y) = \rho^I_{AA} (x,y) + \rho^{II}_{AA*} (x,y)
\eeq
where we find
\beql
\rho^I_{AA} (x,y) = - {1\over 2\pi} {1\over {N^2 (x-y)^2}} {(a^2-xy)\over
[(a^2-x^2)(a^2-y^2)]^{1\over 2}}
\eeql
and
\beql
\rho^{II}_{AA*} (x,y) = { {c^2\mu_1^2} \over {N^2 (2\mu_1\mu_2-c^2)} }
 [ { {\cos (\phi-\theta)-1} \over {2\cos\phi\cos\theta} } \times \nonumber\\ 
{ {2\mu_1^2(x-y)^2-{8\over a^2}(\cos2\phi+\cos2\theta-2\cos(\theta+\phi))
-{8\mu_1\over a}(x-y)(\sin\theta-\sin\phi)} \over { (\mu_1^2(x-y)^2+
{8\over a^2}(1-\cos(\theta-\phi))
-{4\mu_1\over a} (x-y)(\sin\theta-\sin\phi)^2) } } \nonumber \\
+ { {\cos (\phi+\theta)+1} \over {2\cos\phi\cos\theta} } \times \nonumber\\
{ {2\mu_1^2(x-y)^2-{8\over a^2}(\cos2\phi+\cos2\theta+2\cos(\theta+\phi))
-{8\mu_1\over a}(x-y)(\sin\theta-\sin\phi)} \over { (\mu_1^2(x-y)^2+
{8\over a^2}(1+\cos(\theta+\phi))-{4\mu_1\over a} 
(x-y)(\sin\theta-\sin\phi)^2) } } ]
\label{cross}
\eeql
where we have used $\sin \theta = {x\over a}$, $\sin \phi = {y\over a}$
and $a^2={4\over \alpha}$. After some algebra we notice that for 
$\mu_1=\infty$,
\beq
\rho^c_{AA}(x,y)={-1\over {\pi^2 N^2 (x-y)^2}}
\eeq
which is the GOE result and for $\mu_1={2\over N}$,
\beq
\rho^c_{AA}(x,y)={-1\over {2\pi^2 N^2 (x-y)^2} }
\eeq

the GUE result. The expression eq. (\ref{cross}) is relevant for crossover
from the GOE to GUE ensemble. One application of these correlation functions 
is to disordered mesoscopic systems using the transmission matrix formalism
and the other is in the study of unoriented random surfaces.

\section{Conclusions}
\label{conclusions}

In conclusion we have presented two classes of random matrix models. 
One in which there are gaps in the eigenvalue distribution and the other 
in which there are two coupled matrices drawn from the three ensembles
(matrices taken from the same and different ensemble have been considered). 
In each of the models we have derived eigenvalue correlators particularly
density-density correlators. In the first case of gapped matrix models
we have eigenvalue correlators which dependent on $N$ the size of the matrix.
This behaviour persists in the large N thermodynamic limit and for the 
symmetric double-well matrix model parity effects are present. For
the coupled matrix models long range smoothed correlators are found. These
are the parametric and crossover correlators which maybe found in mesoscopic
experiments. Density-density correlators are applicable 
in calculations of conductance fluctuations of mesoscopic conductors. Our
results in these models are valid for all eigenvalues 
near the center as well as the edge of the semi-circle. The behavior near 
the edge of the cut is particularly relevant in studies of transport 
properties of mesoscopic conductors ref. \cite{been}. Thus clever
mesoscopic experiments should be devised which will show the effects
found in both types of these matrix models.  

\section{Acknowledgments}
\label{acknowledgements}

My best wishes and thanks to Professor N. Kumar on the occasion of 
the NKFEST. Thanks to E. Br\'ezin, S. Jain and B. S Shastry for
encouragement and collaborations.    

\noindent
email:ndeo@vsnl.net, ndeo@rri.res.in.

%\section{References}

\end{document}